# NUCLEON-2 Mission for the investigation of heavy cosmic rays' nuclei

**V. Bulatov**[b], **S. Filipov**[b], **D. Karmanov**[a], **I. Kovalev**[a], **A. Kurganov**[a,*], **A. Mansurov**[b], **M. Panasyuk**[a], **A. Panov**[a], **D. Podorozhny**[a], **D. Polkov**[b], **G. Sedov**[a], **P. Tkatchev**[a] **and A. Turundaevskiy**[a]

[a] *SINP MSU,*
  *Skobeltsyn Institute of Nuclear Physics, Moscow State University, Moscow, Russia*
[b] *Gorizont,*
  *Ekaterinburg, Russia*
  *E-mail:* me@sx107.ru

ABSTRACT: The NUCLEON-2 experiment is aimed at the investigation of isotope and charge composition of ions from carbon up to trans-uranium elements in the energy range over about a hundred MeV/N. The concept design of the NUCLEON-2 satellite cosmic ray experiment is presented. The performed simulation and preliminary prototype beam test confirms the isotope resolution algorithms and techniques.

# Contents

## 1. Introduction

### 1.1 Scientific problems connected with the composition of GCR ultra-heavy nuclei

In the current state of space-related fields of physics, including physics of the origin of cosmic rays and astrophysics in general, data on the isotope composition is of considerable interest. This data affects a wide variety of physical problems.

The fluxes of superheavy nuclei in galactic cosmic rays (GCR) are sensitive to the local environment of the Sun, as their free paths are relatively short (~1kpc @ Z>40) before they fragment. Therefore, by measuring the abundances of secondary isotopes one can determine the local diffusion coefficient, knowledge of which is crucial, particularly, for the correct description of positron and electron fraction in cosmic rays. Using many convenient radioisotope clocks among isotopes of superheavy nuclei their age can also be determined. All these measurements are of great importance for the investigation of such exotic and non-standard sources as neutron stars and dark matter.

Supernova explosions in a number of current models may occur in heavy elements-enriched medium formed by earlier supernova explosions and stellar winds, rather than in a standard interstellar medium. This process may lead to a variety of anomalies in isotope and charge composition of GCR that require detailed examination. Models of non-standard cosmic ray acceleration mechanisms in OB stellar associations where the abundance of heavy nuclei in cosmic rays serves as a marker of these mechanisms must also be studied.

From the viewpoint of nucleosynthesis mechanisms examination the data on the isotope composition of cosmic rays is also very important in charge range from Z=40 to 65, as this range contains the double peak of element abundance corresponding to the fast r-process of neutron capture and the slow s-process.

**1.2 Existing data**

A vast majority of data on fluxes of heavy and superheavy nuclei was obtained in four experiments: LDEF, HEAO-C3, SuperTiger and ACE/CRIS.

The LDEF experiment [1] was based on long-term exposure of a solid-state track detector in open space. The experiment was designed to determine the chemical composition of low-energy GCRs in the subactinide ($70 \leq Z \leq 87$) and actinide ($88 \leq Z \leq 103$) regions. It recorded 35 actinide events in the energy range of 1–2 GeV/nucleon and acquired a great many statistics in the subactinide region. The abundances of superheavy elements in GCRs and the Solar System were compared.

The C3 detector at the HEAO-3 astronomical observatory [2] was based on a combination of Cherenkov counters, gas proportional counters, and hodoscopes from multi-wire ionization chambers. The charge spectrum of superheavy nuclei in the charge range of 40 to 62 was determined. A plateau in the abundance of different nuclei was detected in the charge range of 44 to 60. In this region, the experiment recorded from 10 to 30 events for the main even nuclei.

The balloon-borne superTIGER stratospheric experiment [3] was also based on using the Cherenkov technique. Charge distributions in a wide range of charges for superheavy nuclei with charges above $Z = 40$ were obtained for nucleus energies above 2–3 GeV/nucleon, with statistics comparable to those obtained earlier in the HEAO-3-C2 experiment. None of the experiments [1-3] measured the isotopic composition of cosmic-ray nuclei.

Information on isotopes of superheavy nuclei in cosmic rays was obtained in the CRIS experiment on board the ACE spacecraft [4]. The experiment used an approach based on recording the Bragg peak of the complete stopping of nuclei in an array of thin silicon detectors. The spectrometer was carried into orbit in 1997 and is still in operation. The CRIS spectrometer obtained unique charge spectra and isotope compositions of super-heavy nuclei up to $Z = 32$ for energies of several hundreds of MeV/nucleon. It should be emphasized that there are no experimental data on the isotopic composition of nuclei with $Z > 32$, therefore obtaining of such data is extremely important.

**1.3 Proposed mission**

Since no experiments are planned for the investigation of the described scientific problems in the near future and a lack of experimental data on the isotope and charge compositions of cosmic rays at $Z > 32$ for isotope composition and $Z > 40$ for charge composition is present, a new experiment is needed for the continuation of studies in these fields. Considering the drop of GCR flux intensity with increasing charge, an experiment with exposure several magnitudes larger than of the ACE/CRIS one's is needed.

## 2. The NUCLEON-2 Mission

**2.1 Description**

The NUCLEON-2 mission is a proposed satellite experiment project for direct measurement of cosmic rays for the investigation of isotope and charge composition.

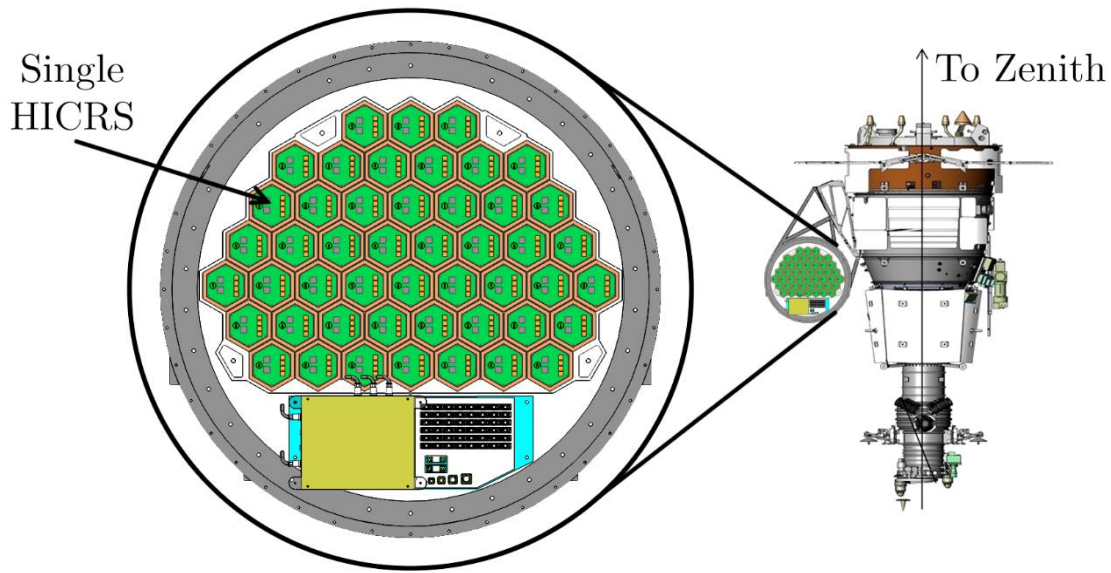

*Figure 1: Possible arrangement of the NUCLEON-2 experiment*

The Z range for charge composition measurement extends from carbon to the stable elements limit; for the isotope composition it is 6-66. The exposition time is not less than 5 years with a planned geometric factor of 0.8 m$^2$sr.

The separation in the experiment uses a modified E-dE telescope technique. This method is based on measuring the particle's total energy deposit E and the energy loss dE/dx in one of the spectrometer's detector. The product of these two variables E(dE/dx) is proportional to the particle mass M and square of the particle charge Z.

The energy range of the spectrometer is 0.1-1 GeV/nucleon, and the energy measurement error does not exceed 0.4%.

To minimize the expenses, it was proposed that the NUCLEON-2 scientific equipment will be launched as a payload on a Russian commercial satellite. The planned orbit is a sun-synchronous low-altitude (400-600km) orbit with 97 degrees' inclination. Mathematical simulations show that at orbits above 400 km the most optimal orientation for a two-sided detector is to orient it horizontally with respect to the Earth. The possible arrangement of the NUCLEON-2 experiment is shown at fig. 1.

## 2.2 Construction of the experiment and Heavy Isotope Cosmic Ray Spectrometer

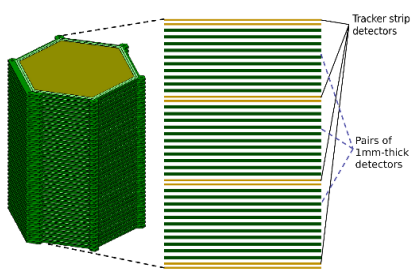

*Figure 2: HICRS construction*

Scientific equipment of the experiment consists of 48 identical Heavy Isotope Cosmic Ray Spectrometers (HICRS). The principles of operation, control, and monitoring would be taken from the NUCLEON scientific equipment as much as possible [5]. The basic characteristics of the NUCLEON-2 scientific equipment are a mass of no more than 295 kg, the device is placed in a cylindrical vessel with dimensions no more than D ~ 1500 H ~ 350 mm and an energy consumption of no higher than 322 W.

Each HICRS is a hexagonal stack of silicon detectors – 32 pairs of 1mm-thick main calorimetric detectors and 8 300um-thick tracker strip detectors with 0.9mm strip pitch (fig. 2), which has two symmetric entrance windows at the top and bottom. The particle charge and mass are determined by measuring the particle's energy deposit in each detector until it stops with a corresponding Bragg peak in the spectrometer using E-dE technique, which is applied multiple times, therefore increasing the accuracy.

**2.3 Expected results**

| Nucleus Z | N | Nucleus Z | N | Nucleus Z | N | Nucleus Z | N |
|---|---|---|---|---|---|---|---|
| 24 | 364500 | 42 | 53 | 60 | 13 | 78 | 18 |
| 25 | 207600 | 43 | 5 | 61 | 3 | 79 | 13 |
| 26 | 4059000 | 44 | 19 | 62 | 11 | 80 | 12 |
| 27 | 18150 | 45 | 23 | 63 | 4 | 81 | 7 |
| 28 | 157900 | 46 | 29 | 64 | 18 | 82 | 13 |
| 29 | 3454 | 47 | 25 | 65 | 4 | 83 | 5 |
| 30 | 2514 | 48 | 32 | 66 | 15 | 84 | 6 |
| 31 | 363 | 49 | 8 | 67 | 2 | 85 | 1 |
| 32 | 468 | 50 | 32 | 68 | 8 | 86 | 2 |
| 33 | 114 | 51 | 9 | 69 | 2 | 87 | 0 |
| 34 | 214 | 52 | 36 | 70 | 9 | 88 | 1 |
| 35 | 136 | 53 | 5 | 71 | 4 | 89 | 0 |
| 36 | 118 | 54 | 22 | 72 | 7 | 90 | 1 |
| 37 | 62 | 55 | 10 | 73 | 4 | 91 | 0 |
| 38 | 176 | 56 | 43 | 74 | 8 | 92 | 4 |
| 39 | 63 | 57 | 4 | 75 | 6 | | |
| 40 | 64 | 58 | 17 | 76 | 12 | | |
| 41 | 37 | 59 | 3 | 77 | 12 | | |

*Table 1: Expected results*

In the low-energy range the nuclei spectra is heavily dependent on solar modulation and geomagnetic cutoff effects. The geomagnetic conditions of the satellite are determined by the orbit inclination on low altitudes. Assuming a circular orbit with 97 degrees inclination and 475km altitude, expected statistics were calculated using flux parametrization [7] while taking solar modulation of the galactic cosmic rays using [8] for the average solar activity. The Stormer geomagnetic cutoff was calculated according to these orbit parameters.

As a first approximation we considered the device to be a silicon chamber with 119 cm diameter an 30cm height. Only the nuclei stopped in the silicon were selected. Thus, the expected experiment statistics from 5 years of exposition were evaluated and are shown in the table 1. The planned geometric factor with this exposition time yields an exposure 20 times larger than of the CRIS ACE experiment.

## 3. The prototype and simulation results

### 3.1 The prototype and CERN test

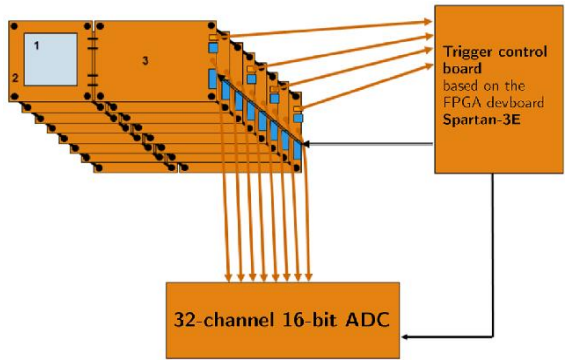

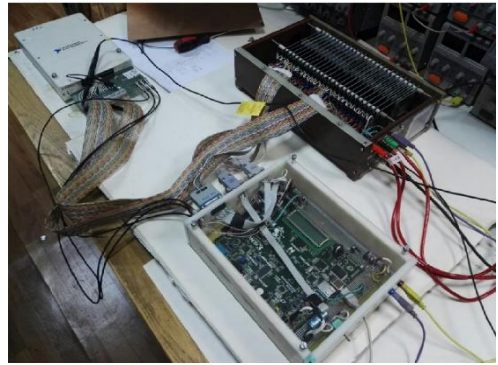

*Figure 4:Prototype block diagram*   *Figure 3:Prototype photo*

A prototype of the HICRS spectrometer was developed. The prototype consists of 13 1mm-thick and 13 300um-thich detectors, giving 26 detectors in total. Each detector is connected throw a preamplifier board to a 32-channel 16-bit ADC, while some of the 300um-thick detectors are connected throw a similar front-end preamplifier to a FPGA trigger controller board based on the Spartan-3E devboard (fig. 3). A photo of the prototype is shown on fig. 4.

The prototype was tested on the SPS accelerator in CERN with a beam of nucleus at 150 GeV/N and Z up to 82 at different A/Z ratios. The results of this test show unchanging 0.17 charge units' resolution throughout the full proposed experiment's Z range from Z=3 up to Z=82. The obtained resolution corresponds well with the simulation performed in Geant 4.10.3. Histogram for the A/Z=2.2 is shown on the fig. 5.

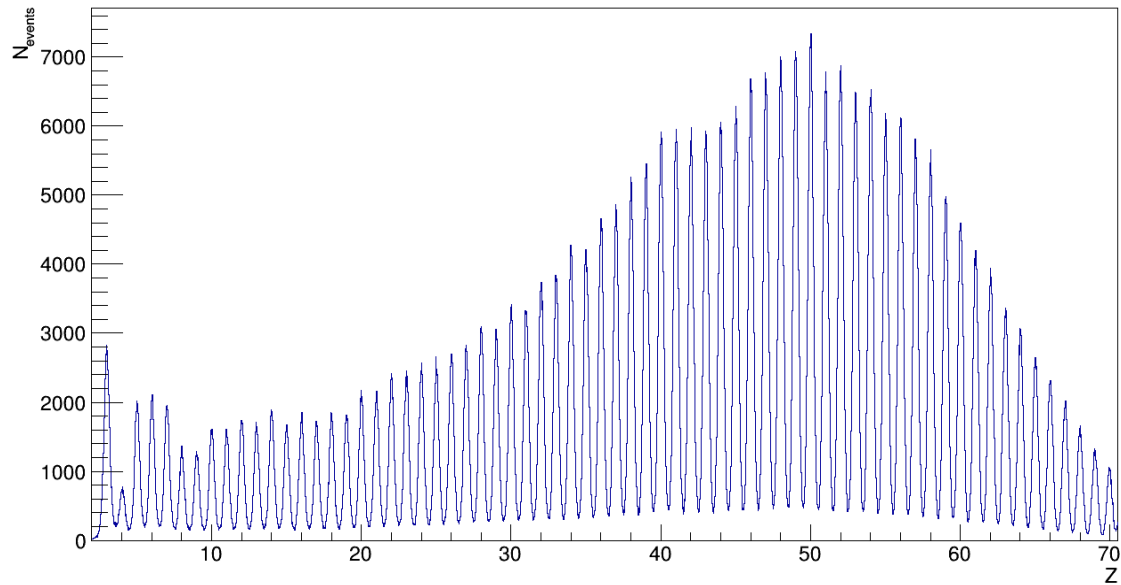

*Figure 5: CERN test results for A/Z=2.2 (Z=2-70)*

### 3.2 Monte-Carlo simulation

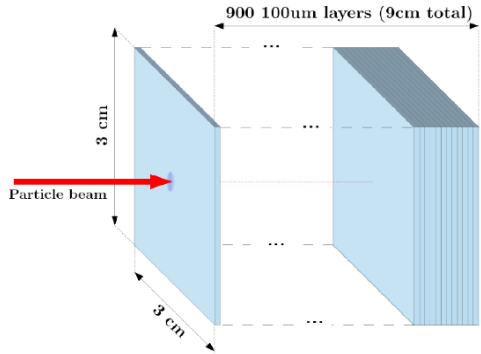

*Figure 6: Model construction*

A Monte-Carlo simulation was performed in GEANT3, GEANT4 and FLUKA codes separately to test the experiment's isotope resolution, optimize the experiment's dimensions and to obtain the acceptable electronics noise level. The model construction consists of 900 100um-thick detectors (fig. 6), which are latterly combined into layers of greater thickness after the simulation. This combination of layers is required to give the possibility of optimizing the experiment's construction without the need of re-simulation. Yet relatively simple, this model still shows basic relations and therefore can be used for the solution of the described simulation tasks.

Simulation data was then analyzed using different processing methods (neural networks and multidimensional analysis based on the maximum likelihood method). Only a weak model dependence was demonstrated by models comparison, and a reliable isotope separation up to $Z = 64$ was obtained.

The separation results for molybdenum and tin isotopes are shown on the fig. 7.

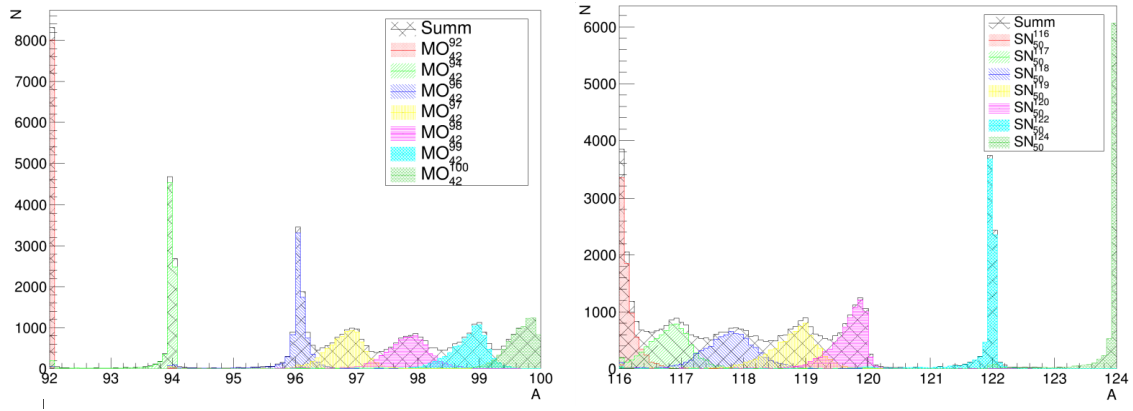

*Figure 7: Monte-Carlo simulation analysis results for molybdenum and tin isotopes at zero noise*

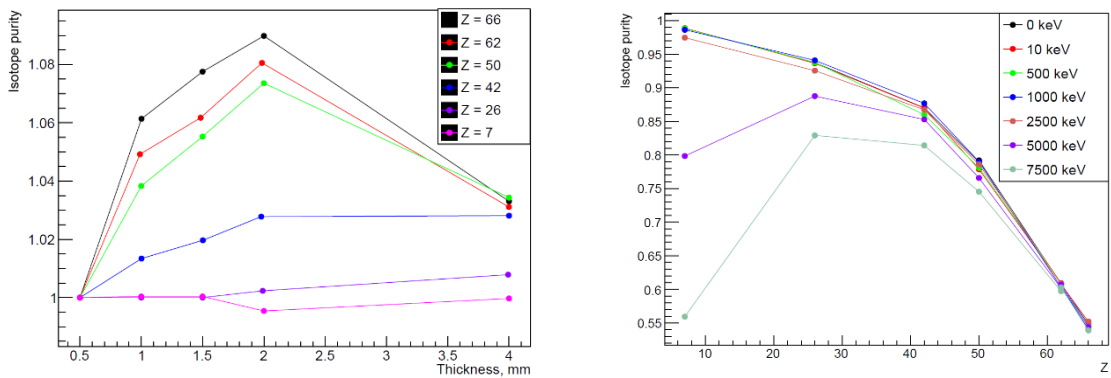

*Figure 8: Isotope separation purity dependence from the detector thickness (on the left) and noise (on the right).*

Using this simulation, the optimal detector thickness and acceptable electronics noise characteristics were obtained. On fig. 8 the isotope separation purity dependence from the detector's thickness and noise is presented. On the left is the isotope separation purity with respect to the isotope separation purity at 0.5mm detector thickness for different charges is shown. A clear peak is seen, especially for high charges, which shows an optimal detector thickness of 2mm.

On the right of fig. 8 is the simulation results on isotope separation purity at different absolute electronic noises and charges. Under 5 MeV noise the isotope separation does not change drastically. It is important to state that the measured electronics noise in the prototype described in 3.1 is under 1 MeV and therefore allows for isotope separation at the full charge range.

## 4. Conclusion

The NUCLEON-2 experiment is a planned experiment for the investigation of isotope composition of superheavy nuclei in GCR. The NUCLEON-2 experiment concept design is presented, as well as the prototype beamtest and simulation results, both of which confirm the validity and operability of the main ideas laid in the detector. The experiment is still in development and will be launched in the year 2020-2022.

## Acknowledgments

We are grateful to ROSCOSMOS State Space Corporation and Russian Academy of Sciences for their continued support of this research.